\documentclass[table]{article}
\usepackage[a4paper, margin=1in]{geometry}
\usepackage{authblk}
\usepackage[T1]{fontenc}
\usepackage{graphicx}
\usepackage{multirow}
\usepackage{caption}
\usepackage{booktabs}
\usepackage[table,xcdraw,svgnames]{xcolor}
\usepackage{tikz,xcolor,hyperref}
\usepackage{subcaption}
\usepackage{rotating}
\usepackage{listings}
\usepackage{pgfplots}
\pgfplotsset{compat=1.18}
\usepackage{pifont}
\usepackage{todonotes}
\definecolor{col1}{HTML}{F67280}
\newcommand{\mybox}[2]{\fcolorbox{#1}{#1!10}{#2}}

\definecolor{dkgreen}{rgb}{0,0.6,0}
\definecolor{gray}{rgb}{0.5,0.5,0.5}
\definecolor{mauve}{rgb}{0.58,0,0.82}

\definecolor{lime}{HTML}{A6CE39}
\providecommand{\keywords}[1]{\textbf{\textit{Index terms---}} #1}
\begin{document}
\title{Outside the Comfort Zone: Analysing LLM Capabilities in Software Vulnerability Detection}
%

% \author{\thanks{Corresponding author.}\inst{1} \and
% Constantinos Patsakis\inst{2,3} \and
% Qiang Hu\inst{4} \and
% Qiang Tang\inst{1} \and
% Fran Casino\inst{2,5,6}}
% \institute{ \\
% \email{yuejun.guo@list.lu,qiang.tang@list.lu} \and
% Information Management Systems Institute, Athena Research Centre (ARC), Artemidos 6, Marousi, Greece \\ 
% \and
% Department of Informatics, University of Piraeus, 80 Karaoli \& Dimitriou str., 18534 Piraeus, Greece \\
% \email{kpatsak@unipi.gr} \and
% Department of Computer Science, The University of Tokyo, Japan \\
% \email{qianghu0515@gmail.com} \and
% Department of Computer Engineering and Mathematics, Rovira i Virgili University, Tarragona, Spain \\ 
% \email{franciscojose.casino@urv.cat}
% }

\author[1]{Yuejun Guo}
\author[2,3]{Constantinos Patsakis}
\author[4]{Qiang Hu}
\author[1]{Qiang Tang}
\author[2,5]{Fran Casino}
\affil[1]{Luxembourg Institute of Science and Technology, Luxembourg}
\affil[2]{Department of Informatics, University of Piraeus, 80 Karaoli \& Dimitriou str., 18534 Piraeus, Greece}
\affil[3]{Information Management Systems Institute of Athena Research Centre, Greece}
\affil[4]{Department of Computer Science, The University of Tokyo, Japan}
\affil[5]{Department of Computer Engineering and Mathematics, Rovira i Virgili University, Tarragona, Spain}
\affil[ ]{yuejun.guo@list.lu, kpatsak@unipi.gr, qiang.tang@list.lu, qianghu0515@gmail.com, franciscojose.casino@urv.cat}

\maketitle % typeset the header of the contribution
\begin{abstract}
The significant increase in software production driven by automation and faster development lifecycles has resulted in a corresponding surge in software vulnerabilities. In parallel, the evolving landscape of software vulnerability detection, highlighting the shift from traditional methods to machine learning and large language models (LLMs), provides massive opportunities at the cost of resource-demanding computations. This paper thoroughly analyses LLMs' capabilities in detecting vulnerabilities within source code by testing models beyond their usual applications to study their potential in cybersecurity tasks. We evaluate the performance of six open-source models that are specifically trained for vulnerability detection against six general-purpose LLMs, three of which were further fine-tuned on a dataset that we compiled. Our dataset, alongside five state-of-the-art benchmark datasets, were used to create a pipeline to leverage a binary classification task, namely classifying code into vulnerable and non-vulnerable. The findings highlight significant variations in classification accuracy across benchmarks, revealing the critical influence of fine-tuning in enhancing the detection capabilities of small LLMs over their larger counterparts, yet only in the specific scenarios in which they were trained. Further experiments and analysis also underscore the issues with current benchmark datasets, particularly around mislabeling and their impact on model training and performance, which raises concerns about the current state of practice. We also discuss the road ahead in the field suggesting strategies for improved model training and dataset curation.
\end{abstract}
\section{Introduction}
\label{sec:intro}
The quest for automation and faster production lifecycles has paved the way for more software solutions. As a result, we have witnessed a massive growth in software over the past few decades, which is mapped to a plethora of digital products and services. Nevertheless, as with all human constructs, software has defects, but in this case, defects are not material. Many of these errors can be identified through the use of the software, as, for instance, the expected functionality is not provided. However, some functionality issues might not be revealed until the software is executed in a specific environment or with parameters that the developer did not expect to handle, either because they are not handled properly, or they are malformed. Of particular interest are security defects, which can expose users to many risks and lead to many hazards since software may handle a lot of sensitive and private information while also being used in critical infrastructures. 

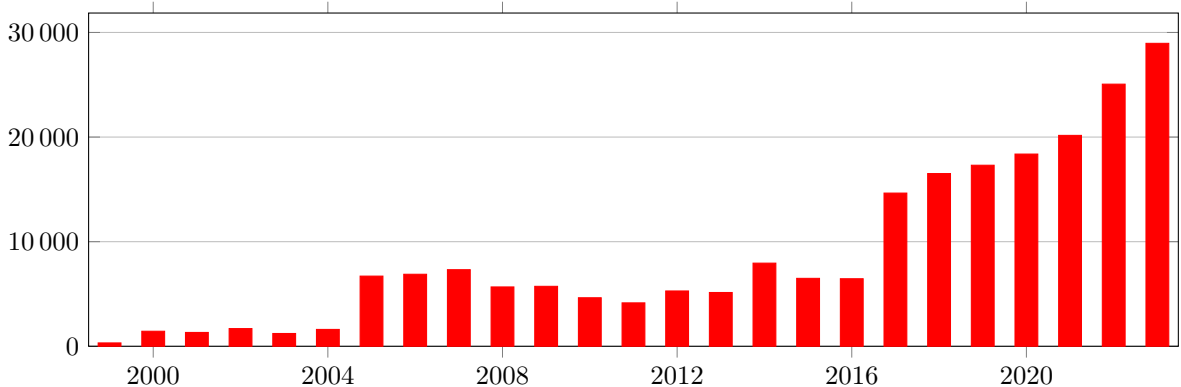
\begin{figure}[ht]

\begin{tikzpicture}
\begin{axis}[
    ybar,
    enlarge x limits=0.02,
    bar width=0.3cm,
    width=\textwidth,
    height=6cm,
    ymin=0,
 ymajorgrids=true,
    scaled  ticks=false,
    xtick={2000,2004,2008,2012,2016,2020},
    symbolic x coords={1999, 2000, 2001, 2002, 2003, 2004, 2005, 2006, 2007, 2008, 2009, 2010, 2011, 2012, 2013, 2014, 2015, 2016, 2017, 2018, 2019, 2020, 2021, 2022, 2023, 2024},
    yticklabel style={
 /pgf/number format/fixed,
 /pgf/number format/1000 sep={\,}
    },
]
\addplot [red,fill=red] coordinates {
    (2023,28961)
    (2022,25059)
    (2021,20161)
    (2020,18375)
    (2019,17308)
    (2018,16512)
    (2017,14645)
    (2016,6457)
    (2015,6494)
    (2014,7948)
    (2013,5142)
    (2012,5288)
    (2011,4150)
    (2010,4639)
    (2009,5732)
    (2008,5673)
    (2007,7322)
    (2006,6885)
    (2005,6708)
    (2004,1612)
    (2003,1223)
    (2002,1691)
    (2001,1323)
    (2000,1438)
    (1999,321)
};
\end{axis}
\end{tikzpicture}
\caption{Number of CVEs by year. Source: \protect\url{https://www.cve.org/About/Metrics}}
\label{fig:cve_by_year}
\end{figure}

Moreover, there has been a continuous increase in the number of reported software vulnerabilities. As illustrated in Fig.~\ref{fig:cve_by_year}, the number of vulnerabilities has quadrupled in the last decade. We argue that this can be attributed to the parallel introduction of the General Data Protection Regulation (GDPR) and the issuance of the Presidential Policy Directive 41 (PPD-41) which pushed private and public organisations to report cyber security incidents. Notably, we have a doubling of reported vulnerabilities in 2017, just the next year of their introduction.

In parallel, the exponential increase in data generation, leading to the creation of vast datasets, has been crucial for the increasing capabilities and complexity of artificial intelligence (AI) and machine learning (ML) in real-world applications. Advances in computational technology (e.g., Graphical Processing Units or GPUs and Tensor Processing Units or TPUs) have enabled unforeseen processing capabilities, enabling the use of resource-demanding models. The introduction of large language models (LLMs) has revolutionised how machines understand, interpret, and generate human language and their democratisation by vast communities behind their use and adoption, such as Hugging Face~\cite{huggingface}, has raised the attention not just of the research community but society as a whole. As more advanced human-machine interactions arise, LLMs are acquiring broader capabilities and application scenarios due to their ability to improve and adapt to new data and contexts.  

As one would expect, researchers and the industry quickly stepped in to harness the new capabilities of ML and LLMs to timely and accurately identify vulnerabilities. This is a major shift as previously one would resort to traditional monolithic solutions like regular expressions~\cite{semgreprules} to identify vulnerable code. The inherent flexibility in coding styles allows developers to express themselves differently, resulting in low accuracy and precision for traditional detection tools, as patterns can be easily bypassed or falsely triggered by non-vulnerable code. The major development in this new era is that models specifically trained in vulnerable and secure code can timely and more accurately identify vulnerable code before it reaches production. This is even more relevant for LLMs as they are able to generate, understand, and summarise code quite efficiently.

In this paper, we perform a thorough analysis of LLMs' capabilities in detecting source code vulnerabilities. Nevertheless, we try to push LLMs beyond their comfort zone and understand the possible gaps and pitfalls in this process. We conduct a series of comprehensive experiments with different benchmark datasets and observe how they perform. This broad variation leads us to propose the use of fine-tuning strategies to leverage low resource-demanding LLMs in the task of software vulnerability detection. Our strategies illustrate that while fine-tuning can enable these LLMs to outperform larger ones in particular contexts, it may lead to a loss of generalisation ability. Indeed, we observe a lot of variation in their capacities when changing the underlying dataset. Finally, we assess the benchmark datasets using state-of-the-art tools that are used in the industry. The latter enables us to provide some fruitful discussion and a strategy to improve the training and accuracy of LLMs in the future. Our approach relies on several research questions pertinent to code vulnerability detection, as described in Table~\ref{tab:resq}.

\begin{table*}[hbt!]
\renewcommand{\arraystretch}{1.1}
\setlength{\tabcolsep}{3pt} % Default value: 6pt
 \centering
  \caption{Summary of research questions and the corresponding sections devoted to answering them.}
 \scriptsize
 \rowcolors{2}{gray!15}{white}
 \begin{tabular}{p{0.35\linewidth}p{0.45\linewidth}c}
\toprule \textbf{Research Question} & \multicolumn{1}{c}{\textbf{Objective}} & \textbf{Discussion}\\
 \midrule
\textbf{RQ1}. Which methods are currently used for software vulnerability detection?    &  The objective is to summarise the current state of the art and approaches towards identifying vulnerabilities in source code. & Section~\ref{sec:related} \\
\textbf{RQ2} Can base LLMs detect vulnerabilities in source code?  &  The objective is to evaluate the capabilities of general-purpose LLMs and their performance towards software vulnerability analysis. & Sections~\ref{sec:related},~\ref{sec:exp}, \ref{sec:discussion} \\
\textbf{RQ3} Is fine-tuning an enabling strategy to improve the trade-off between computational resources and detection accuracy?  &  The objective is to provide insight towards the use of fine-tuning to improve the performance of base LLMs to a level in which they outperform larger models.  & Section~\ref{sec:discussion} \\
\textbf{RQ4} How robust are the analysed vulnerability detection models?   &  The objective is to assess whether the models can generalise across different benchmark datasets. & Sections~\ref{sec:exp}, \ref{sec:discussion}\\
\textbf{RQ5} Are curation and labelling methodologies employed on existing datasets robust enough for training LLMs and ensuring their desired functionality?  &  We try to assess existing datasets using industry tools to determine how well they are labelled and whether they provide the necessary information for proper training. & Sections~\ref{sec:exp}, \ref{sec:discussion}\\
\textbf{RQ6} Given the analysis and outcomes provided in this paper, what are the next steps towards software vulnerability detection?   &  The objective is to analyse the lessons learned in this paper and provide a view regarding desired functionalities and capabilities of generative AI towards source code analysis. & Sections~\ref{sec:discussion}, \ref{sec:conclusions}\\
 \bottomrule
 \end{tabular}
 \scriptsize
 \label{tab:resq}%
\end{table*}%

The remainder of this work is structured as follows. In the next section, we provide an overview of the related work regarding code vulnerability detection. Then, in Section~\ref{sec:method}, we introduce the reader to our methodology and code vulnerability analysis pipeline. Section~\ref{sec:exp} introduces the datasets used and the experimental setup. Next, in Section~\ref{sec:discussion} we report the outcomes and provide a detailed discussion. Finally, the paper concludes in Section \ref{sec:conclusions}, recalling our research findings and proposing ideas for future work.

\section{Related Work}
\label{sec:related}
We review related work from the perspective of three areas: static application security testing (SAST), task-specific deep learning (DL) models, and large language models (LLMs) for vulnerability detection.

\subsection{SAST-Based Vulnerability Detection}
SAST tools typically utilize rule-based~\cite{Lee2006rule} and signature-based~\cite{Senanayake2023android} methods to scan the source code for vulnerabilities, which require predefined rules or patterns indicative of known vulnerabilities. The scanning technique varies between different tools, and the popular ones are pattern matching~\cite{viega2002token}, symbolic execution~\cite{Sebastian2020symbolic}, and data-flow analysis~\cite{Sampaio2016explore}. Croft~\cite{Croft2021empirical} conducted an empirical study involving three SAST tools (Flawfinder~\cite{flawfinder}, Cppcheck~\cite{cppcheck}, and RATs~\cite{RATS}) selected from the tool lists provided by NIST~\cite{nist} and OWASP~\cite{owasp}, demonstrating that ML-based approaches provide better overall performance for detection and assessment. Other widely used SAST tools are Semgrep~\cite{semgrep}, SNYK~\cite{Snyk} and Sonarqube~\cite{sonarqube}. Although effective in certain contexts, these tools are usually time-consuming to develop considering the required domain knowledge on security weaknesses and may need to be adjusted to identify novel or previously unknown vulnerabilities. 

\subsection{Task-Specific DL Models for Vulnerability Detection}
Compared to traditional static analysers that often require manual feature engineering by security experts, AI automates the analysis and can function at different granularities (e.g., file, function, and program slice). Various AI models, especially DL models, have been developed and put in use. Typically, a DL model is initialized with random parameters and then trained on a set of labelled data containing both vulnerable and non-vulnerable code samples. This type of model, also known as task-specific model, is specifically designed to detect vulnerabilities within codebases. VulDeePecker~\cite{li2018vuldeepecker} proposed by Li \emph{et al.} is a very early work that adopted DL techniques to automatically identify vulnerabilities in source code. Specifically, VulDeePecker utilizes BLSTM networks to learn vulnerable information and outperforms pattern-based and code similarity-based methods. Later, Zhou \emph{et al.} proposed Devign~\cite{zhou2019devign}, a Graph Neural Networks~(GNNs) based method to detect vulnerabilities. The key component of Devign involves leveraging GNNs to learn from code with semantic representations, such as Abstract Syntax Tree~(AST). The dataset provided by Devign has been widely studied in the vulnerability detection field. More recently, Chen \emph{et al.} built a new dataset~\cite{diversevul2023} that contains 18,945 vulnerable functions for the performance evaluation of vulnerability detection models. The experimental results demonstrated that existing vulnerability detection models perform poorly in their dataset. The well-known empirical study~\cite{chakraborty2021deep} showed that existing methods cannot generalize to real-world vulnerability prediction and can be improved by using a proper pipeline combining the best practices in each process, such as data collection and model design. Besides, some surveys~\cite {Senanayake2023android,lin2020software} comprehensively reviewed the literature that lies in the direction of DL-based vulnerability detection. 

\subsection{LLM-Based Vulnerability Detection}
The advent of LLMs is changing the vulnerability detection paradigm. Rather than being explicitly trained on labelled vulnerable and non-vulnerable source code, LLMs are pre-trained on vast amounts of data from various sources, such as online blogs, books, and code repositories. During pre-training, these models learn to capture statistical patterns and semantic meanings in the data. When fine-tuned on vulnerable and non-vulnerable source code, these models leverage their pre-trained knowledge to identify vulnerability patterns and often outperform task-specific models. Based on the RoBERTa~\cite{liu2019roberta} architecture tailored for text-based tasks, Microsoft developed CodeBERT~\cite{codebert2020} and Hanif~\emph{et al.}~\cite{Hanif2022vulberta} introduced VulBERTa specifically for source code analysis. Both models have been widely used and fine-tuned using different datasets for vulnerability detection~\cite{msr2024,reveal2024,d2a2024,draper2024,guo2024imbalance}. Ribeiro \emph{et al.}~\cite{ribeiro2023gpt} explored the application of GPT-3 for automatically fixing type errors in OCaml code, with a particular focus on addressing general bugs and providing some discussion regarding vulnerabilities. Noever \emph{et al.}~\cite{noever2023can} tried to utilize GPT-4~\cite{openai2024gpt4} to find and fix vulnerabilities and found that GPT-4 can correct programs and reduce 90\% vulnerabilities. Li \emph{et al.}~\cite{li2023assisting} focused on using GPT-3.5 and GPT-4 to identify vulnerabilities by providing code and context. Both models demonstrated the ability to identify vulnerable code and provide detailed information on the detected vulnerabilities. Charalambous \emph{et al.}~\cite{charalambous2023new} combined LLMs and formal verification techniques to identify vulnerabilities. The proposed method can detect and repair 80\% of vulnerable code. Zheng \emph{et al.}~\cite{zheng2023towards} reviewed the use of LLMs for software engineering tasks, including vulnerability detection, and found that LLMs do not perform well on this task according to the results reported in existing work. More recently, Zhou \emph{et al.}~\cite{zhou2024large} reported their emerging results of LLMs for vulnerability detection and showed that GPT-3.5 has competitive performance with fine-tuned CodeBERT~\cite{codebert2020} and GPT-4 significantly outperforms fine-tuned CodeBERT in this task. Lastly, Zhou \emph{et al.}~\cite{zhou2024large} surveyed 36 works related to LLMs for vulnerability detection and repair and discussed the challenges and research directions in this task.

Contrary to previous work, our study further investigates the impact of fine-tuning on detection performance. Additionally, we study the labelling issue in existing datasets and its impact on model performance.

\section{Methodology}
\label{sec:method}
\begin{figure*}[ht]
\centering
\includegraphics[width=\textwidth]{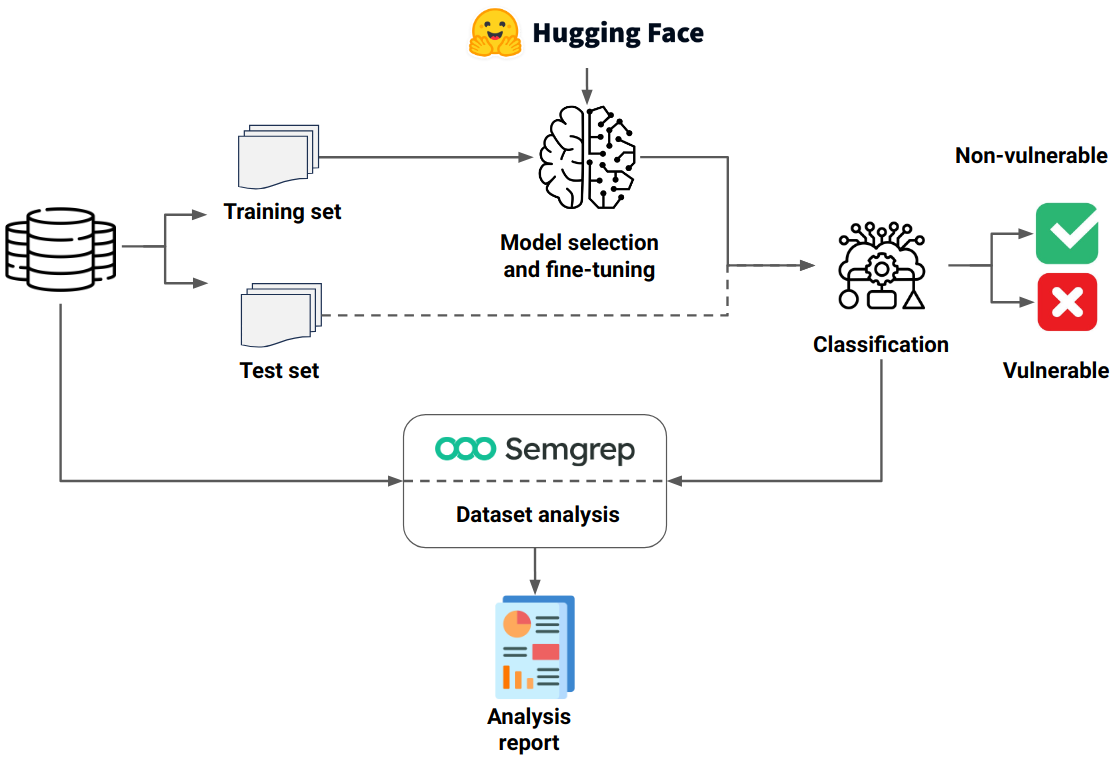}
\caption{Overview of our methodology.} \label{fig:method}
\end{figure*}

Our approach to analysing the potential vulnerabilities in the source code is based on a comprehensive methodology designed to maximise efficiency and accuracy, as seen in Figure~\ref{fig:method}. The process begins with collecting code samples in the form of curated datasets. We may assume that these datasets are correctly labelled, yet this is not true according to our experiments and the state of the art~\cite{ding2024vulnerability,diversevul2023}. The latter motivated the incorporation of a dataset quality test using commercial/industry tools. In this regard, we selected Semgrep~\cite{semgrep}, yet any set of similar tools could be used. Therefore, we apply Semgrep to assess the quality of the datasets in conjunction with the classification outcomes, as discussed in Section~\ref{sec:discussion}. To assess whether a code sample is vulnerable, we apply two strategies. First, we select a dataset, split it into training and testing sets, and use the training set to fine-tune a model. Otherwise, we may use a dataset to directly test a model (i.e., by using the whole dataset as a test set or just a split of it). In all cases, we select models from the Hugging Face Hub~\cite{huggingface}, yet other pools could be used. From this hub, we selected a subset of models of two categories: models explicitly trained for code vulnerability detection and models designed for general-purpose use. Note that further selection can be made based on performance and hardware requirements; thus, we enriched our model selection based on such criteria. Therefore, given a dataset and a model, we perform an optional fine-tuning procedure and a binary classification task, which assesses whether a code sample is vulnerable or non-vulnerable. 

According to our criteria, we selected six models from Hugging Face that were already trained specifically for code vulnerability detection:
\begin{itemize}
    \item The first five models, developed by Claudio, are VulBERTa-MLP-ReVeal~\cite{reveal2024}, \emph{VulBERTa-MLP-D2A}~\cite{d2a2024}, \emph{VulBERTa-MLP-Draper}~\cite{draper2024}, \emph{VulBERTa-MLP-MVD}~\cite{mvd2024}, and \emph{VulBERTa-MLP-VulDeePecker}~\cite{VulDeePecker2024}. These models share the same architecture and are trained on the ReVeal~\cite{Reveal2022}, D2A~\cite{d2a2021zheng}, Draper~\cite{russell2018automated}, muVuldeepecker~\cite{miu2019zou}, and Vuldeepecker~\cite{vuldeepecker2018li} datasets, respectively. VulBERTa-MLP is a fully-connected layer with 768 neurons and one output layer 2 neurons. 
    \item The sixth model is Codebert\_fine\_tuned\_detect\_vulnerability\_on\_MSR (CodeBERT\_finetuned\_MSR)~\cite{msr2024}, which has been fine-tuned on the CodeBERT model using the MSR dataset~\cite{Fan2020MSR}.
\end{itemize}

Further to these trained models, we also selected six LLMs to test their code vulnerability detection capabilities, such as CodeLLama, Mistral, and OpenAI's GPT-4 since they are often used as baselines in the state of the art: 

\begin{itemize}
    \item CodeBERT-base~\cite{codebert2020}, developed by Microsoft, is a pre-trained model for general-purpose code understanding. It has been trained on the CodeSearchNet dataset~\cite{husain2020codesearchnet} that includes 2.4 million functions. 
    \item Mistral-7b-base~\cite{jiang2023mistral}, developed by the Mistral AI team, is a pre-trained generative text model with 7.3 billion parameters. It is a decoder-only model.
    \item Mixtral-8x7b-base~\cite{jiang2024mixtral}, also developed by the Mistral AI team, is a high-quality sparse mixture of experts model (SMoE) with open weights. This model is pre-trained on data extracted from the open Web and has 46.7 billion parameters. Mixtral-8x7b shares the same architecture as Mistral-7b, but the main difference is that the feedforward block of Mixtral-8x7b picks from a set of 8 distinct groups of parameters. 
    \item CodeLlama~\cite{roziere2024codellama} is a family of code-specialized LLMs. It supports many of the most popular languages that are used today, including Python, C++, Java, PHP, Typescript (Javascript), C\#, and Bash. In this work, we use the \emph{CodeLlama-7b-base} and \emph{CodeLlama-13b-base} with 7 and 13 billion parameters for comparison. 
    \item GPT-4-base~\cite{openai2024gpt4}, developed by OpenAI, is a large multimodal model (accepting image and text inputs, emitting text outputs). GPT-4 is available on ChatGPT Plus and as an API for developers.
\end{itemize}

Except for GPT-4-base, all other base models are publicly available on Hugging Face. To adapt CodeBERT-base for vulnerability detection, we construct a custom model using the architecture of \verb|RobertaForSequenceClassification| from the Hugging Face Transformers~\cite{2020transformers}. The model is configured to be the same as \verb|microsoft/codebert-base| and the default maximum sequence length (512 tokens) of CodeBERT to encode code samples. Regarding Mistral and CodeLLama models, we build them utilising the architecture of \verb|AutoModelForSequenceClass| \verb|-ification| with the \verb|num_labels| set to 2. Additionally, we apply the 4-bit quantization through \verb|BitsAndBytes| for efficient evaluation as depicted in Fig.~\ref{fig:tune}(a). 

As described in our pipeline, we fine-tuned a subset of these base models using a dataset crafted by us, later described in Section \ref{sec:datasets}. The fine-tuned models are described as follows:

\begin{itemize}
    \item CodeBERT-fine-tuned shares the same architecture as CodeBERT-base. The fine-tuning is for 50 epochs, and the model with the minimum loss on the test set is saved for testing.
    \item Regarding CodeLlama-7b-fine-tuned and Mistral-7b-fine-tuned, we use the PEFT LoRA~\cite{hu2022lora} which stands for Parameter Efficient Fine Tuning (PEFT) using Low-Rank Adaptation (LoRA) method for efficient fine-tuning with 3 epochs, as depicted in Fig.~\ref{fig:tune}(b). Note that the configuration \verb|task_typetask_type=| \verb|TaskType.SEQ_CLS| is essential for specifying the task type as a sequence classification.
\end{itemize}

\begin{figure*}[th]
    \centering
   \begin{subfigure}[h]{0.5\textwidth}
 \centering
\begin{lstlisting}
#4-bit quantization is applied through BitsAndBytesConfig: 
load_in_4bit=True
bnb_4bit_use_double_quant=True
bnb_4bit_quant_type="nf4"
bnb_4bit_compute_dtype=torch.bfloat16 
\end{lstlisting}
 \caption{BitsAndBytes configuration.}
    \end{subfigure}~
    \begin{subfigure}[h]{0.5\textwidth}
 \centering
\begin{lstlisting}
#LoRA is configured via LoraConfig as follows: 
task_type=TaskType.SEQ_CLS
r=32
lora_alpha=64
bias="none" 
lora_dropout=0.05
\end{lstlisting}
 \caption{LoRA configuration.}
    \end{subfigure}~
   \caption{Detail of the fine-tuning configuration of the selected LLMs.}
    \label{fig:tune}
\end{figure*}

\section{Experiments}
\label{sec:exp}
\subsection{Prompt Engineering and Hardware Setup}
Since crafting prompts that guarantee the expected responses and comprehension by the employed models requires brute-force trial-and-error experimentation, the proper ones were selected after several iterations. In the case of all models, when available, the temperature was set to zero to allow for reproducibility and reduce the hallucinations in local models.

We used OpenAI's GPT-4 (\texttt{gpt-4-0613}) via the API offered by OpenAI in our experiments. Fig.~\ref{fig:prompt_tasks} shows the prompt to obtain the detection results. Note that since GPT-4 is a paid service, and the charges for its API are based on token consumption, due to budget constraints, we tested it only in our dataset and the Lin2017 dataset (see Section \ref{sec:datasets}.

\begin{figure*}[htpb]
    \centering 
 \begin{tabular}{|p{\textwidth}|}
   \hline
   \cellcolor{col1!10}\textbf{System:} You are a senior developer doing security code auditing.\\   
   \hline \textbf{User:} Check the following code for vulnerabilities. If you find one, return only 1, otherwise return 0. Suppress all other output. The code is the following :  \texttt{\textasciigrave\textasciigrave\textasciigrave}\mybox{col1}{CODE}\texttt{\textasciigrave\textasciigrave\textasciigrave} \\
  % \hline \cellcolor{col2!10}\textbf{Assistant:} "PERSON"\\
   \hline
 \end{tabular}
    \caption{Structure of the task prompts used in OpenAI's GPT.}
    \label{fig:prompt_tasks}
\end{figure*}

The rest of the experiments were conducted on a high-performance computer (HPC) cluster and each cluster node runs a 2.20GHZ Intel Xeon Silver 4210 Processor with an NVIDIA Tesla V100-PCIE-32GB GPU. Models are trained and tested using the PyTorch 2.0.1 framework with CUDA 12.0. 

\subsection{Datasets}
\label{sec:datasets}
We used six datasets in our experiments\footnote{All the used datasets are unified and publicly available on Zenodo~\cite{yuejun_2024_10975439}.}. The details of the datasets can be seen in Table~\ref{tbl:dataset_data}. First, we created a balanced dataset that includes 13,532 code functions written in C. The 6,766 vulnerable functions were manually collected from projects on GitHub that have registered CVEs in NVD from 2002 to 2023. The 6,766 non-vulnerable code functions are extracted from the DiverseVul dataset~\cite{diversevul2023} to increase the code diversity. The entire dataset is divided randomly, with 80\% allocated for fine-tuning (i.e., we use this dataset to fine-tune a subset of models as described in Section~\ref{sec:method}) and 20\% for testing purposes, while ensuring a balanced representation of both vulnerable and non-vulnerable functions. Next, we selected five open-source datasets yet, this time only for testing purposes. For Devign~\cite{zhou2019devign}, Lin2017~\cite{Lin2017discovery}, Choi2017~\cite{Choi2017end}, and LineVul~\cite{linevul2022fu}, we collected all their available data to construct the datasets for testing. While for PrimeVul~\cite{ding2024vulnerability}, we only used its test set to ensure a fair and direct comparison with the results outlined in the original paper~\cite{ding2024vulnerability}, where the data was originally sourced and evaluated.

\begin{table}[th]
    \centering
    \caption{Detail of datasets and their composition. }
    \label{tbl:dataset_data}
    \resizebox{0.75\textwidth}{!}{%
\begin{tabular}{llccc}
\toprule
\textbf{Ref.} & \textbf{Dataset} & \textbf{\#Vulnerable} & \textbf{\#Non-vulnerable} & \textbf{\#Total} \\ \hline
- & Our dataset & 6,766 & 6,766 & 13,532 \\
\cite{zhou2019devign} & Devign & 12,460 & 14,858 & 27,318 \\
\cite{Lin2017discovery} & Lin2017 & 44 & 577 & 621 \\
\cite{Choi2017end} & Choi2017 & 7,054 & 6,946 & 14,000 \\
\cite{linevul2022fu} & LineVul & 1,055 & 17,809 & 18,864 \\
\cite{ding2024vulnerability} & PrimeVul & 695 & 25,213 & 25,908 \\
\bottomrule
\end{tabular}
}
\end{table}

\section{Results and Discussion}
\label{sec:discussion}

\begin{sidewaystable}
    \centering
    \setlength{\tabcolsep}{3pt} % Default value: 6pt
    \renewcommand{\arraystretch}{1.3}
    \tiny
    \caption{For each dataset and model, we report the P (precision), R (recall) and F1-score per class (i.e., vulnerable, non-vulnerable) to ease their interpretation.}
    \resizebox{\textwidth}{!}{%
\begin{tabular}{ll
>{\columncolor[HTML]{E0E0E0}}c 
>{\columncolor[HTML]{E0E0E0}}c 
>{\columncolor[HTML]{E0E0E0}}c 
>{\columncolor[HTML]{E0E0E0}}c 
>{\columncolor[HTML]{E0E0E0}}c 
>{\columncolor[HTML]{E0E0E0}}c cccccc
>{\columncolor[HTML]{E0E0E0}}c 
>{\columncolor[HTML]{E0E0E0}}c 
>{\columncolor[HTML]{E0E0E0}}c 
>{\columncolor[HTML]{E0E0E0}}c 
>{\columncolor[HTML]{E0E0E0}}c 
>{\columncolor[HTML]{E0E0E0}}c }
\toprule 
&  & \multicolumn{6}{c}{\cellcolor[HTML]{E0E0E0}\textbf{Choi2017}} & \multicolumn{6}{c}{\textbf{LineVul}} & \multicolumn{6}{c}{\cellcolor[HTML]{E0E0E0}\textbf{PrimeVul}} \\
& & \multicolumn{3}{c}{\cellcolor[HTML]{E0E0E0}\textbf{Vulnerable }} & \multicolumn{3}{c}{\cellcolor[HTML]{E0E0E0}\textbf{Non-vulnerable }} & \multicolumn{3}{c}{\textbf{Vulnerable }} & \multicolumn{3}{c}{\textbf{Non-vulnerable }} & \multicolumn{3}{c}{\cellcolor[HTML]{E0E0E0}\textbf{Vulnerable }} & \multicolumn{3}{c}{\cellcolor[HTML]{E0E0E0}\textbf{Non-vulnerable }} \\
\multirow{-2}{*}{\textbf{Ref.}} & \multirow{-2}{*}{ \textbf{Model}} & \multicolumn{1}{l}{\cellcolor[HTML]{E0E0E0}\textbf{P}} & \multicolumn{1}{l}{\cellcolor[HTML]{E0E0E0}\textbf{R}} & \multicolumn{1}{l}{\cellcolor[HTML]{E0E0E0}\textbf{F1}} & \multicolumn{1}{l}{\cellcolor[HTML]{E0E0E0}\textbf{P}} & \multicolumn{1}{l}{\cellcolor[HTML]{E0E0E0}\textbf{R}} & \multicolumn{1}{l}{\cellcolor[HTML]{E0E0E0}\textbf{F1}} & \multicolumn{1}{l}{\textbf{P}} & \multicolumn{1}{l}{\textbf{R}} & \multicolumn{1}{l}{\textbf{F1}} & \multicolumn{1}{l}{\textbf{P}} & \multicolumn{1}{l}{\textbf{R}} & \multicolumn{1}{l}{\textbf{F1}} & \multicolumn{1}{l}{\cellcolor[HTML]{E0E0E0}\textbf{P}} & \multicolumn{1}{l}{\cellcolor[HTML]{E0E0E0}\textbf{R}} & \multicolumn{1}{l}{\cellcolor[HTML]{E0E0E0}\textbf{F1}} & \multicolumn{1}{l}{\cellcolor[HTML]{E0E0E0}\textbf{P}} & \multicolumn{1}{l}{\cellcolor[HTML]{E0E0E0}\textbf{R}} & \multicolumn{1}{l}{\cellcolor[HTML]{E0E0E0}\textbf{F1}} \\
\hline
\cite{reveal2024} & VulBERTa-MLP-ReVeal & 0 & 0 & 0 & 49.61    & 100   & 66.32 & 16.27  & 16.87   & 16.57 & 95.06  & 94.86   & 94.96 & 2.10 & 40.43 & 3.99  & 98.20    & 89.55 & 93.67 \\
\cite{d2a2024} & VulBERTa-MLP-D2A    & 50.39    & 100   & 67.01 & 0 & 0 & 0 & 5.93   & 41.33   & 10.37 & 94.62  & 61.15   & 74.29 & 2.10 & 33.24 & 3.96  & 96.89    & 57.36 & 72.06 \\
\cite{draper2024} & VulBERTa-MLP-Draper & 0 & 0 & 0 & 49.61    & 100   & 66.32 & 0 & 0   & 0   & 94.41  & 100 & 97.12 & 0 & 0 & 0 & 97.32    & 100   & 98.64 \\
\cite{mvd2024} & VulBERTa-MLP-MVD    & 57.96    & 1.29  & 2.52  & 49.70    & 99.05 & 66.19 & 9.85   & 7.01    & 8.19  & 94.58  & 96.20   & 95.38 & 5.32 & 11.80 & 7.33  & 97.48    & 94.21 & 95.82 \\
\cite{VulDeePecker2024} & VulBERTa-MLP-VulDeePecker & 0 & 0 & 0 & 49.61    & 100   & 66.32 & 14.62  & 2.37    & 4.08 & 94.49  & 99.18   & 96.78 & 4.78 & 2.45  & 3.24  & 97.35    & 98.66 & 98    \\
\cite{msr2024} & CodeBERT\_finetuned\_MSR & 50.28    & 15.21 & 23.36 & 49.60    & 84.73 & 62.57 & 9.29   & 5.50    & 6.91  & 94.53  & 96.82   & 95.66 & 6.30 & 63.31 & 11.47 & 98.65    & 74.06 & 84.61 \\
\cite{codebert2020} & CodeBERT-base & 50.39 & 100 & 67.01 & 0 & 0 & 0 & 5.59 & 100 & 10.59 & 0 & 0 & 0 & 2.68 & 100 & 5.23 & 0 & 0 & 0 \\
\cite{jiang2023mistral} & Mistral-7b-base     & 50.38    & 99.99 & 67    & 0 & 0 & 0 & 4.86   & 50.62   & 8.86  & 93.38  & 41.26   & 57.23 & 2.33 & 75.97 & 4.53  & 94.91    & 12.34 & 21.84 \\
\cite{jiang2024mixtral} & Mixtral-8x7b-base   & 50.89    & 29.17 & 37.09 & 49.82    & 71.41 & 58.69 & 6.04   & 44.83   & 10.65 & 94.73  & 58.69   & 72.48 & 5.88 & 0.43  & 0.80  & 97.32    & 99.81 & 98.55 \\
\cite{roziere2024codellama} & CodeLlama-7b-base   & 44.99    & 13.69 & 20.99 & 48.64    & 83    & 61.33 & 5.34   & 49  & 9.64  & 94.15  & 48.59   & 64.10 & 3.20 & 40.58 & 5.93  & 97.58    & 66.13 & 78.84 \\
\cite{roziere2024codellama} & CodeLlama-13b-base    & 49.04    & 70.37 & 57.80 & 46.09    & 25.73 & 33.02 & 2.45   & 29  & 4.51  & 88.22  & 31.50   & 46.42 & 4.16 & 12.95 & 6.29  & 97.45    & 91.77 & 94.52 \\
\cite{openai2024gpt4} & GPT-4-base & - & - & - & - & - & - & - & -   & -   & - & -   & -   & - & - & - & - & - & - \\
\hline
Our & CodeBERT-fine-tuned & 50.32    & 97.87 & 66.47 & 46.43    & 1.87  & 3.60  & 10.78  & 63.13   & 18.42 & 96.93  & 69.06   & 80.66 & 5.30 & 85.90 & 9.99  & 99.33    & 57.71 & 73.01 \\
Our & Mistral-7b-fine-tuned   & 50.18    & 72.63 & 59.35 & 49.06    & 26.78 & 34.65 & 6.35   & 86.16   & 11.84 & 96.80  & 24.78   & 39.46 & 3.04 & 89.21 & 5.88  & 98.64    & 21.54 & 35.36 \\
Our & CodeLlama-7b-fine-tuned & 50.39    & 100   & 67.01 & 0 & 0 & 0 & 5.96   & 94.88   & 11.21 & 97.37  & 11.24   & 20.16 & 2.80 & 95.40 & 5.44  & 98.56    & 8.71  & 16.01 \\
%Our  & Voting 1 out of n   & 50.39    & 100   & 67.01 & - & - & - & 5.88   & 96.87   & 11.08 & - & -   & -   & 2.79 & 97.55 & 5.42  & - & - & - \\
%Our  & Majority voting     & 50.39    & 100   & 67.01 & - & - & - & 5.88   & 96.87   & 11.08 & - & -   & -   & 2.79 & 97.55 & 5.42  & - & - & -   \\
\bottomrule
\end{tabular}
}
\newline
%\vspace*{0.5 cm}
%\newline
\resizebox{\textwidth}{!}{%
\begin{tabular}{ll
>{\columncolor[HTML]{E0E0E0}}c 
>{\columncolor[HTML]{E0E0E0}}c 
>{\columncolor[HTML]{E0E0E0}}c 
>{\columncolor[HTML]{E0E0E0}}c 
>{\columncolor[HTML]{E0E0E0}}c 
>{\columncolor[HTML]{E0E0E0}}c cccccc
>{\columncolor[HTML]{E0E0E0}}c 
>{\columncolor[HTML]{E0E0E0}}c 
>{\columncolor[HTML]{E0E0E0}}c 
>{\columncolor[HTML]{E0E0E0}}c 
>{\columncolor[HTML]{E0E0E0}}c 
>{\columncolor[HTML]{E0E0E0}}c }
\toprule 
   &   & \multicolumn{6}{c}{\cellcolor[HTML]{E0E0E0}\textbf{Our Dataset}}   & \multicolumn{6}{c}{\textbf{Devign}} & \multicolumn{6}{c}{\cellcolor[HTML]{E0E0E0}\textbf{Lin2017}}   \\
   &   & \multicolumn{3}{c}{\cellcolor[HTML]{E0E0E0}\textbf{Vulnerable }}    & \multicolumn{3}{c}{\cellcolor[HTML]{E0E0E0}\textbf{Non-vulnerable }}  & \multicolumn{3}{c}{\textbf{Vulnerable }}  & \multicolumn{3}{c}{\textbf{Non-vulnerable }}   & \multicolumn{3}{c}{\cellcolor[HTML]{E0E0E0}\textbf{Vulnerable }}    & \multicolumn{3}{c}{\cellcolor[HTML]{E0E0E0}\textbf{Non-vulnerable }}  \\
\multirow{-2}{*}{\textbf{Ref.}} & \multirow{-2}{*}{\textbf{Model}} & \multicolumn{1}{l}{\cellcolor[HTML]{E0E0E0}\textbf{P}} & \multicolumn{1}{l}{\cellcolor[HTML]{E0E0E0}\textbf{R}} & \multicolumn{1}{l}{\cellcolor[HTML]{E0E0E0}\textbf{F1}} & \multicolumn{1}{l}{\cellcolor[HTML]{E0E0E0}\textbf{P}} & \multicolumn{1}{l}{\cellcolor[HTML]{E0E0E0}\textbf{R}} & \multicolumn{1}{l}{\cellcolor[HTML]{E0E0E0}\textbf{F1}} & \multicolumn{1}{l}{\textbf{P}} & \multicolumn{1}{l}{\textbf{R}} & \multicolumn{1}{l}{\textbf{F1}} & \multicolumn{1}{l}{\textbf{P}} & \multicolumn{1}{l}{\textbf{R}} & \multicolumn{1}{l}{\textbf{F1}} & \multicolumn{1}{l}{\cellcolor[HTML]{E0E0E0}\textbf{P}} & \multicolumn{1}{l}{\cellcolor[HTML]{E0E0E0}\textbf{R}} & \multicolumn{1}{l}{\cellcolor[HTML]{E0E0E0}\textbf{F1}} & \multicolumn{1}{l}{\cellcolor[HTML]{E0E0E0}\textbf{P}} & \multicolumn{1}{l}{\cellcolor[HTML]{E0E0E0}\textbf{R}} & \multicolumn{1}{l}{\cellcolor[HTML]{E0E0E0}\textbf{F1}} \\
\hline
\cite{reveal2024}  & VulBERTa-MLP-ReVeal & 78.31    & 17.07 & 28.03 & 53.46    & 95.27 & 68.49 & 51.25  & 8.38    & 14.40 & \cellcolor[HTML]{FFFFFF}54.84  & \cellcolor[HTML]{FFFFFF}93.32 & \cellcolor[HTML]{FFFFFF}69.08    & 26.55    & 68.18 & 38.22 & 97.24    & 85.62 & 91.06 \\
\cite{d2a2024} & VulBERTa-MLP-D2A    & 48.06    & 51.37 & 49.67 & 36.27    & 44.49 & 39.96 & 46.93  & 52.36   & 49.49 & \cellcolor[HTML]{FFFFFF}55.75  & \cellcolor[HTML]{FFFFFF}50.34 & \cellcolor[HTML]{FFFFFF}52.91    & 7..44    & 5.23  & 13.03 & 93.27    & 50.43 & 65.47 \\
\cite{draper2024}  & VulBERTa-MLP-Draper & 0 & 0 & 0 & 50   & 100   & 66.67 & 0 & 0   & 0   & \cellcolor[HTML]{FFFFFF}54.39  & 100 & \cellcolor[HTML]{FFFFFF}70.46    & 0 & 0 & 0 & 92.91    & 100   & 96.33 \\
\cite{mvd2024} & VulBERTa-MLP-MVD    & 70.59    & 6.21  & 11.41 & 50.95    & 97.41 & 66.90 & 47.60  & 4.21    & 7.74  & \cellcolor[HTML]{FFFFFF}54.47  & \cellcolor[HTML]{FFFFFF}96.11 & \cellcolor[HTML]{FFFFFF}69.53    & 6.90 & 4.55  & 5.48  & 92.91    & 95.32 & 94.10 \\
\cite{VulDeePecker2024}  & VulBERTa-MLP-VulDeePecker & 74.23    & 5.32  & 9.93  & 50.90    & 98.15 & 67.04 & 56.72  & 3.35    & 6.33  & \cellcolor[HTML]{FFFFFF}54.70  & \cellcolor[HTML]{FFFFFF}97.85 & \cellcolor[HTML]{FFFFFF}70.17    & 25   & 2.27  & 4.17  & 93.03    & 99.48 & 96.15 \\
\cite{msr2024} & CodeBERT\_finetuned\_MSR & 86.39    & 9.39  & 16.93 & 52.09    & 98.52 & 68.15 & 50.82  & 6.50    & 11.53 & \cellcolor[HTML]{FFFFFF}54.71  & \cellcolor[HTML]{FFFFFF}94.72 & \cellcolor[HTML]{FFFFFF}69.36    & 15.04    & 90.91 & 25.81 & 98.87    & 60.83 & 75.32 \\
\cite{codebert2020}    & CodeBERT-base  & 50   & 100   & 66.67 & 0 & 0 & 0 & 45.61  & 100 & 62.65 & 0 & 0   & 0   & 7.09 & 100   & 13.23 & 0 & 0 & 0 \\
\cite{jiang2023mistral}  & Mistral-7b-base     & 49.09    & 93.50 & 64.38 & 31.78    & 3.03 & 5.53  & 45.57  & 99.37   & 62.48 & 46.26  & 0.46    & 0.91  & 28.07    & 36.36 & 31.68 & 95.04    & 92.89 & 93.95 \\
\cite{jiang2024mixtral}   & Mixtral-8x7b-base   & 83.33    & 0.37  & 0.74  & 50.07    & 99.93 & 66.72 & 42.15  & 43.13   & 42.64 & 51.36  & 50.36   & 50.86 & 2.86 & 36.36 & 5.31  & 54.84    & 5.89  & 10.64 \\
\cite{roziere2024codellama}     & CodeLlama-7b-base   & 51.57    & 60.61 & 55.73 & 52.24    & 43.09 & 47.23 & 46.19  & 82.40   & 59.19 & 48.76  & 13.38   & 21  & 42.86    & 34.09 & 37.97 & 95.05    & 96.53 & 95.79 \\
\cite{roziere2024codellama}     & CodeLlama-13b-base    & 48.76    & 92.98 & 63.97 & 24.60    & 2.29  & 4.19  & 45.25  & 92.95   & 60.86 & 48.95  & 5.67    & 10.16 & 5.65 & 47.73 & 10.10 & 90.76    & 39.17 & 54.72 \\
\cite{openai2024gpt4}  & GPT-4-base & 55.91    & 95.42 & 70.51 & 84.38    & 24.76 & 38.29 & - & -   & -   & - & -   & -   & 8.63 & 100    & 15.88 & 100 & 19.24    &32.27 \\
\hline
Our  & CodeBERT-fine-tuned & 70.22    & 74.94 & 72.50 & 73.12    & 68.14 & 70.54 & 47.98  & 63.08   & 54.51 & 57.94  & 42.65   & 49.13 & 16.60    & 95.45 & 28.28 & 99.46    & 63.43 & 77.46 \\
Our  & Mistral-7b-fine-tuned   & 88.70    & 85.88 & 87.27 & 86.32    & 89.06 & 87.67 & 45.87  & 92.17   & 61.25 & 57.20  & 8.77    & 15.21 & 7.63 & 100   & 14.17 & 100  & 7.63  & 14.17 \\
Our  & CodeLlama-7b-fine-tuned & 97.97    & 96.30 & 97.13 & 96.37    & 98    & 97.18 & 45.91  & 97.17   & 62.36 & 62.12  & 4.14    & 7.76  & 7.09 & 100   & 13.23 & 100  & 0.52  & 1.03 \\
%Our  & Voting 1 out of n   & 73.10    & 98.23 & 83.82 & - & - & - & 45.67  & 99.02   & 62.51 & - & -   & -   & 7.09 & 100   & 13.23 & - & - & - \\
%Our  & Majority voting     & 73.10    & 98.23 & 83.82 & - & - & - & 45.67  & 99.02   & 62.51 & - & -   & -   & 7.09 & 100   & 13.23 & - & - & -    \\   
\bottomrule
\end{tabular}
}
\label{tab:outcomes_exps}
\end{sidewaystable}

Table~\ref{tab:outcomes_exps} shows the outcomes for each model and dataset. In all experiments, we employ three widely-used metrics~\cite{zhou2019devign}, namely precision, recall, and F1-score (F1), to evaluate the detection performance. We computed such metrics per class as they showcase specific behaviours related to the models' performance. %Moreover, we considered a voting scheme in which the identification of vulnerable code when using our fine-tuned models was evaluated. More concretely, if at least one of the models considers code as vulnerable, we flag it as such in the ``Voting 1 out of n'' method. In the case of the ``Majority voting'' method, we consider a code as vulnerable only if at least two out of three methods identify a code sample as vulnerable. 

Regarding the Choi2017 dataset, we observe that models reporting high recall values for the vulnerable class do not perform well for the non-vulnerable class and vice-versa. Practically, this means that some models classify all code as vulnerable and thus cannot classify the code with qualitative criteria. The best-performing models are Mixtral-8x7b-base and Mistral-7b-fine-tuned with F1-score values close to 47\%, when combining both classes (i.e., we average the F1-score outcomes to provide an indicative value to be used as reference, as the unbalanced nature of the datasets is already collected in the values per class). In the case of LineVul, we observe similar behaviour for some models (e.g., CodeBERT-base shows the same behaviour in all datasets tested), yet we observe a remarkable identification of non-vulnerable code for most models, while vulnerable code is poorly identified. Overall, VulBERTa-based variations obtain the highest F1-score considering both classes, closely followed by CodeBERT-fine-tuned. The outcomes of PrimeVul dataset are similar to those obtained in LineVul, yet this time, CodeLlama-based models and Mixtral-8x7b-base perform similarly to VulBERTa-based variations, obtaining around 50\% of F1-score considering both classes. We also observed a similar behaviour in the case of Lin2017, with the difference that best scoring models obtained scores above 60\%, as in the case of CodeLlama-7b-base and VulBERTa-MLP-ReVeal. Since LineVul, PrimeVul and Lin2017 are not balanced, the tests showcase the capability of the models to identify non-vulnerable code, as the number of samples is higher. The latter also reinforces the relevance of reporting the outcomes per class, as unbalanced classes could hide underperforming issues~\cite{casino2021intercepting}. Choi2017, Devign and our dataset are balanced regarding samples per class. In general, models obtain between 30\% and 40\% of F1-score considering the average of both classes, being CodeBERT-fine-tuned and VulBERTa-MLP-D2A the best-performing ones, with values around 51\%. Finally, the outcomes obtained on our dataset showcase the contextual nature of fine-tuning. In this regard, while state-of-the-art models perform similarly to the rest of the datasets (i.e., with averaged F1-score values raging between 30\% and 50\% considering both classes), all our fine-tuned models achieve values over 70\%, with CodeLlama-7b-fine-tuned achieving a remarkable 97\%. 

The previously discussed outcomes raised our curiosity, as they highlighted incongruencies and alarmingly low accuracy in the code vulnerability detection task. We wanted to delve into this and performed another experiment to explore the quality of the datasets. As reported in the state of the art,~\cite{ding2024vulnerability,diversevul2023} several widely-used datasets present mislabelling issues due to, e.g., the use of automated tools for annotation, or treating code as fixed after a commit even though the vulnerability was not solved. Thus, further to merely using the datasets, we used Semgrep~\cite{semgrep} to scan code snippets using the command-line interface - CLI with Semgrep public rules~\cite{semgreprules}. While there are other tools such as Snyk~\cite{Snyk} and SonarQube~\cite{sonarqube} that are often used to detect source code vulnerabilities, none of them could be used in our experiments. The reason is that both these tools operate on projects and not code segments, as in the case of these datasets. To determine whether there is a vulnerability, they need full access to the code to assess the imported libraries, dependencies, etc. However, this is not the case for Semgrep. Quite interestingly, Semgrep faced many issues in scanning the code snippets of the datasets. In fact, for each dataset, it reported scanning issues in the form of a message: \texttt{Partially scanned: X files only partially analyzed due to parsing or internal Semgrep errors}, with X varying in each dataset (the \#Issues column in Table~\ref{tbl:dataset_comp}). As noted in Table~\ref{tbl:dataset_comp}, even in these cases, Semgrep identified a very small fragment of the vulnerabilities that each dataset contains. Therefore, we assert that the information provided in all datasets may not be sufficient for existing industry tools to reliably determine whether the code snippets are vulnerable. The latter showcases the previously stated concerns regarding dataset labelling, which requires further analysis to ensure that models are not trained with erroneous data or data that can create conflicts. 

\begin{table}[th]
    \setlength{\tabcolsep}{4pt} % Default value: 6pt
    \centering
    \caption{Detail of the Semgrep detection result. \#Vulnerable from labeled vulnerable/non-vulnerable code: the number of vulnerable code detected by Semgrep from the labeled vulnerable/non-vulnerable data, respectively. \#Issues: number of data that happened the partially scanned issue.}
    \label{tbl:dataset_comp}
    \resizebox{\textwidth}{!}{%
    \begin{tabular}{llcc
>{\columncolor[HTML]{E0E0E0}}c 
>{\columncolor[HTML]{E0E0E0}}c
>{\columncolor[HTML]{E0E0E0}}c
}
\hline
\textbf{} & \textbf{} & \multicolumn{2}{c}{\textbf{Labeling in the dataset}} & \multicolumn{3}{c}{\cellcolor[HTML]{E0E0E0} \textbf{Semgrep detection}} \\
\textbf{Ref.} & \textbf{Dataset} & \textbf{\#Vulnerable} & \textbf{\#Non-vulnerable} & \textbf{\begin{tabular}[c]{@{}c@{}}\#Vulnerable from \\ labeled vulnerable code\end{tabular}} & \textbf{\begin{tabular}[c]{@{}c@{}}\#Vulnerable from \\ labeled non-vulnerable code\end{tabular}} & \textbf{\#Issues} \\ \hline
- & Our dataset & 1353 & 1353 & 75 & 3 & 745 \\
\cite{zhou2019devign} & Devign & 12460 & 14858 & 166 & 169 & 3708 \\
\cite{Lin2017discovery} & Lin2017 & 44 & 577 & 26 & 0 & 185 \\
\cite{Choi2017end} & Choi2017 & 7054 & 6946 & 0 & 0 & 10000 \\
\cite{linevul2022fu} & LineVul & 1055 & 17809 & 104 & 0 & 10842 \\
\cite{ding2024vulnerability} & PrimeVul & 695 & 25213 & 41 & 161 & 11975 \\ \hline
    \end{tabular}
    }
\end{table}

Simultaneously, this raises another important question. If such tools are not able to detect vulnerabilities in such code fragments, how sure are we that the provided information is enough for LLMs to find vulnerabilities? For instance, tools like Semgrep use rules that describe string patterns\footnote{\url{https://semgrep.dev/docs/writing-rules/rule-ideas/}} to find vulnerable code. Nevertheless, the extent of failure of such tools in identifying vulnerabilities in the code fragments of the datasets can potentially signify that what the LLM understands from its training is very limited or not precise enough, making it mark all code as vulnerable. While we expect the LLM's tokenizer to accurately segment code into tokens, the task of identifying the roles (e.g. variable names and functions) to understand that, e.g., passing unprocessed user input to a function can lead to a code injection attack goes beyond its capability. However, Semgrep and similar tools already have the rule for that and fail to detect the vulnerability. While one could consider that the tokenizer of Semgrep is not good enough, since this is a well-established tool, we opt to attribute such failures to lack of proper contextual understanding. Indeed, this could just justify our research findings and the failure of LLMs to accurately find vulnerabilities when tested in different datasets. We argue that LLMs generate broad rules based on their training tokens, which can incorrectly mark code fragments as vulnerable due to their limited ability to discern the code context. Even worse, Semgrep identifies vulnerabilities in code that was labelled secure in several datasets, raising even more questions about the quality of the datasets. 
Even if the detection results are false positives, the fact that they are detected by such a tool implies that the LLMs could be wrongfully trained and fail to identify the proper patterns. 

\section{Conclusions}
\label{sec:conclusions}

The advent of generative AI tools and the sophistication of software production have enhanced the lifecycle and robustness of digital products and services. Nevertheless, the analysis of the current state of practice reveals that we are only beginning to scratch the surface regarding LLMs' capabilities. Thus, significant efforts must be devoted to realising accurate and efficient automated code vulnerability detection. The research questions posed in Section \ref{sec:intro} summarise the main aim of our research, namely providing a comprehensive analysis of the state of practice in code vulnerability detection analysis through the use of AI, its main challenges and elaborating a fruitful discussion on this particular matter. We discuss them in order as follows:

\begin{quote}
    \textit{\textbf{RQ1:}  Which methods are currently used for source code vulnerability detection?  }
\end{quote}

To provide enough background to discuss the current state of the art, we provide an extensive analysis of related work, including traditional SAST-based, task-specific DL models, and LLM-based vulnerability detection. As discussed in Section \ref{sec:related}, LLMs are gaining momentum and therefore it is crucial to study their potential.

\begin{quote}
    \textit{\textbf{RQ2:}  Can base LLMs detect vulnerabilities in source code?}
\end{quote}
Given the analysis of the state of the art and the experiments performed in this paper, the answer to that question is unclear. One could argue that LLMs can effectively detect vulnerabilities in source code, yet their accuracy is particularly tied to their training data, which generally performs primarily on the patterns included in training data. Furthermore, larger models exhibit more stable accuracy across datasets yet still do not achieve remarkable outcomes.

\begin{quote}
    \textit{\textbf{RQ3:}   Is fine-tuning an enabling strategy to improve the trade-off between computational resources and detection accuracy? }
\end{quote}
Given the outcomes analysed in Section \ref{sec:discussion}, fine-tuning allows low resource-demanding models to outperform larger ones in specific contexts. Despite having fewer parameters than commercial models, local LLMs can be fine-tuned to optimise their performance in specific tasks as their weights are made publicly available, which we will explore in future work. The latter includes exploring smaller LLMs (e.g., through quantisation \cite{dettmers2024qlora} and number of parameters) to provide resource-efficient solutions, fostering the adoption of LLMs in constrained environments.  Nevertheless, this entails several constraints, such as the generalisation issues discussed in RQ4.  

\begin{quote}
    \textit{\textbf{RQ4:}  How robust are the analysed detection models?}
\end{quote}

As seen in Section \ref{sec:discussion}, the extent of the application context is closely tied to the training data since models usually do not generalise well when exposed to different testing environments. The latter requires a specific analysis of the benchmarks, as modifications can derive unexpected model behaviour and classification errors. Moreover, data curation is a parallel issue, as discussed in RQ5. 

\begin{quote}
    \textit{\textbf{RQ5:}  Are curation and labelling methodologies employed on existing datasets robust enough for training LLMs and ensuring their desired functionality? }
\end{quote}

As highlighted in the state of the art  and according to our dataset analysis experiments with Semgrep, there are concerning issues regarding the labelling of datasets. Issues such as the length of the code sample and the use of automated strategies with flaws create contradictory judgements about the samples. While this could only mean the inability to evaluate models properly, in the case of LLMs this incurs further fundamental issues, as they are trained on these datasets, thus corrupting the entire functionality, as in, e.g., poisoning attacks. 

\begin{quote}
    \textit{\textbf{RQ6:}  Given the analysis and outcomes provided in this paper, what are the next steps towards software vulnerability detection ?}
\end{quote}

This work provides a clear insight into the current state of practice and critical aspects that should be improved towards the reliability of LLMs and similar models. In this regard, our future research paths are aligned with our outcomes and focus on producing quality datasets. The latter can be done by establishing a sound methodology to guarantee that they can be used in software development, security, and operations cycles (e.g., by ensuring formatting and length, curation, and providing data related to the CWEs to allow precise and reliable evaluation). In parallel, we aim to delve into how LLMs acquire knowledge, e.g., by fine-tuning processes, to avoid overfitting and optimising their generalisation capabilities. Finally, aspects related to explainability and pedigree, namely which datasets were used to train and create models, are essential to ensure their robustness and avoid biased evaluations.

\section*{Acknowledgements} 
This work was supported by the European Commission under the Horizon Europe Programme, as part of the projects CyberSecPro (\url{https://www.cybersecpro-project.eu}) (Grant Agreement no. 101083594) and LAZARUS (\url{https://lazarus-he.eu/}) (Grant Agreement no. 101070303). This work was partially supported by Ministerio de Ciencia, Innovación y Universidades, Gobierno de España (Agencia Estatal de Investigación, Fondo Europeo de Desarrollo Regional -FEDER-, European Union) under the research grant PID2021-127409OB-C33 CONDOR. Fran Casino was supported by the Government of Catalonia with the Beatriu de Pinós programme (Grant No. 2020 BP 00035), and by AGAUR with the project ASCLEPIUS (2021SGR-00111).

The content of this article does not reflect the official opinion of the European Union. Responsibility for the information and views expressed therein lies entirely with the authors.

\end{document}